\documentclass[prl,superscriptaddress,twocolumn,showpacs]{revtex4}

\usepackage{amsfonts,amssymb,amsmath}
\usepackage[]{graphics,graphicx,epsfig}
\usepackage{amsthm}

\begin{document}

\title{Bosonic bunching reveals strong contextual behaviour}

\author{Pawe\l\ Kurzy\'nski}
\email{cqtpkk@nus.edu.sg}
\affiliation{Centre for Quantum Technologies,
National University of Singapore, 3 Science Drive 2, 117543 Singapore,
Singapore}
\affiliation{Faculty of Physics, Adam Mickiewicz University,
Umultowska 85, 61-614 Pozna\'{n}, Poland}

\author{Akihito Soeda}
\email{cqtas@nus.edu.sg}
\affiliation{Centre for Quantum Technologies,
National University of Singapore, 3 Science Drive 2, 117543 Singapore,
Singapore}

\author{Jayne Thompson}
\email{cqttjed@nus.edu.sg}
\affiliation{Centre for Quantum Technologies,
National University of Singapore, 3 Science Drive 2, 117543 Singapore,
Singapore}

\author{Dagomir Kaszlikowski}
\email{phykd@nus.edu.sg}
\affiliation{Centre for Quantum Technologies,
National University of Singapore, 3 Science Drive 2, 117543 Singapore,
Singapore}
\affiliation{Department of Physics,
National University of Singapore, 3 Science Drive 2, 117543 Singapore,
Singapore}




\begin{abstract}
We show that violation of Klyachko-Can-Binicioglu-Shumovsky [Phys. Rev. Lett. {\bf 101}, 020403 (2008)] pentagram-like inequality can exceed $\sqrt{5}$ provided that exclusive events do not have to be comeasurable and that one uses bosonic systems which exhibit bunching effects. We also show that in this case one can find three pairwise exclusive events whose sum of probabilities is $3/2$.
\end{abstract}

\maketitle

Contextuality is defined as a dependence of the measurment outcome on the choice of which other measurements are simultaneously performed. It was proven by Kochen and Specker \cite{KS} that any quantum system of dimension greater than two is contextual. Recently Klyachko-Can-Binicioglu-Shumovsky (KCBS) proved that for five cyclically exclusive events (that is, at most one of events $i$ and $i+1$, for $i=1,\dots,5$ modulo 5, can happen) quantum mechanics does not allow joint probability distributions in accord with a non-contextual hidden variable model. KCBS derived an inequality for probabilities of these five events and showed that the sum of their probabilities cannot exceed $2$ for any non-contextual hidden variable theory. It was also shown \cite{KCBS,Pent,C2} that in quantum mechanics the sum of probabilities for five cyclically orthogonal projective measurements can reach at most $\sqrt{5}$. 

In this note we present a physical  system where the sum of probabilities for five cyclically exclusive events exceeds the quantum bound of $\sqrt{5}$ and reachs the maximal value of $5/2$ for the 5 event KCBS-like inequality. In addition we also analyze the previously considered  test \cite{CSW} with three pairwise exclusive events that was believed to not exhibit contextuality in quantum systems and show that in fact quantum system can exhibit maximal contextuality. Note that these so called quantum bounds can hypothetically be violated if the no-disturbance principle is not obeyed  \cite{G, us}.  However, in this paper we stick to this principle and still show the violation.  Finally we discuss the cause of violation.

We focus on a particular realisation of bosonic bunching with photons, which was demonstrated experimentally by Hong-Ou-Mandel \cite{HOM}. However, we would like to point out that the following argument would apply to arbitrary bosons.
\begin{figure}[t]
\begin{center}
\includegraphics[scale=0.1]{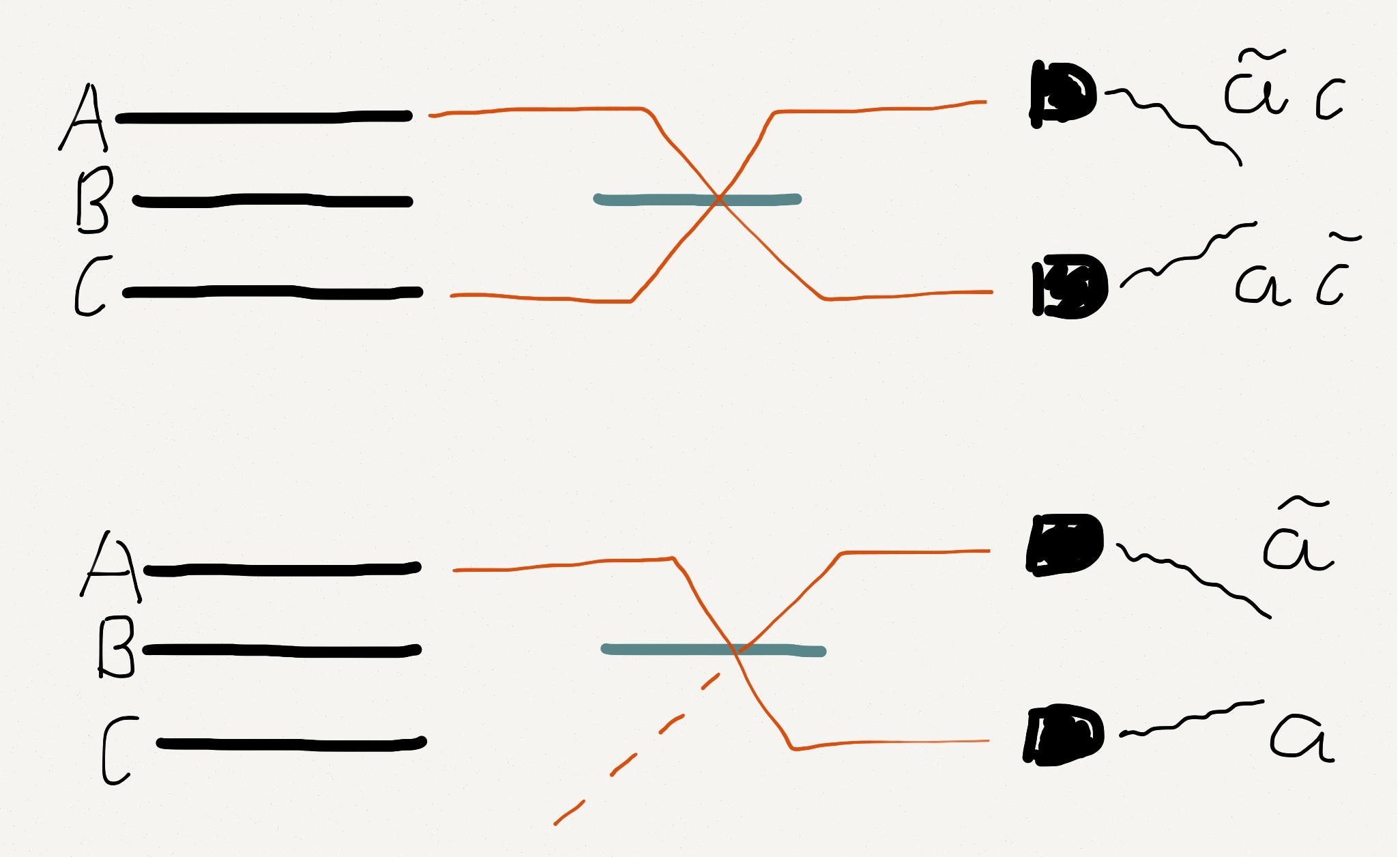}
\end{center}
\caption{\label{f1} Schematic picture representing the setup and possible outcomes for two-photon and single-photon events.}
\end{figure}
Let us consider three optical fibers $A$,$B$, $C$ and a single $50-50$ beamsplitter (BS). One can connect at most two out of the three fibers to the BS's input ports (see Fig. \ref{f1}). In each fiber there is a single photon that is identical apart from its fiber assignment. The rules of the game are simple. You choose either one or two arbitrary fibers, connect them to the BS's input ports and observe what happens at the BS's output ports. Every photon can be either reflected or transmitted thru the BS. We assume that this property of photons does not depend on which input port they enter thru. For example, the conditional probability of the event that the photon from fiber A is reflected and the photon from C transmitted given that $A$ and $C$ were connected to the BS is denoted as $p(\tilde{a}c|AC)$.  Here the tilde symbol is used to signify the reflected case and non-tilde letters for the transmitted. Due to the fact that photons are indistinguishable one has $$p(x\tilde{y}|XY)=p(x\tilde{z}|XZ),$$ that is, the probability of outcomes behind the BS cannot depend on whether the photon came from $Y$ or $Z$. The same holds for all the other events $p(xy|XY)$, $p(\tilde{x}y|XY)$, $p(\tilde{x}\tilde{y}|XY)$. Note that events $x\tilde{y}$ and $\tilde{x}y$ can be distinguished becasue both the inputs and outputs of the BS can be labeled. In addition, the no-disturbance assumption holds for all events, that is, $$\sum_y p(xy|XY)=\sum_z p(xz|XZ)=p(x|X).$$

Imagine that it is possible to assign outcomes to all observable events before the actual experiment happens. As we said before, we assume that the behavior of photons does not depend on which of the BS's port they enter thru. Moreover, assume that this can be done in a non-contextual way, that is, the property that photon X is reflected or transimtted does not depend on the input of the other port. 

\begin{figure}[t]
\begin{center}
\includegraphics[scale=0.12]{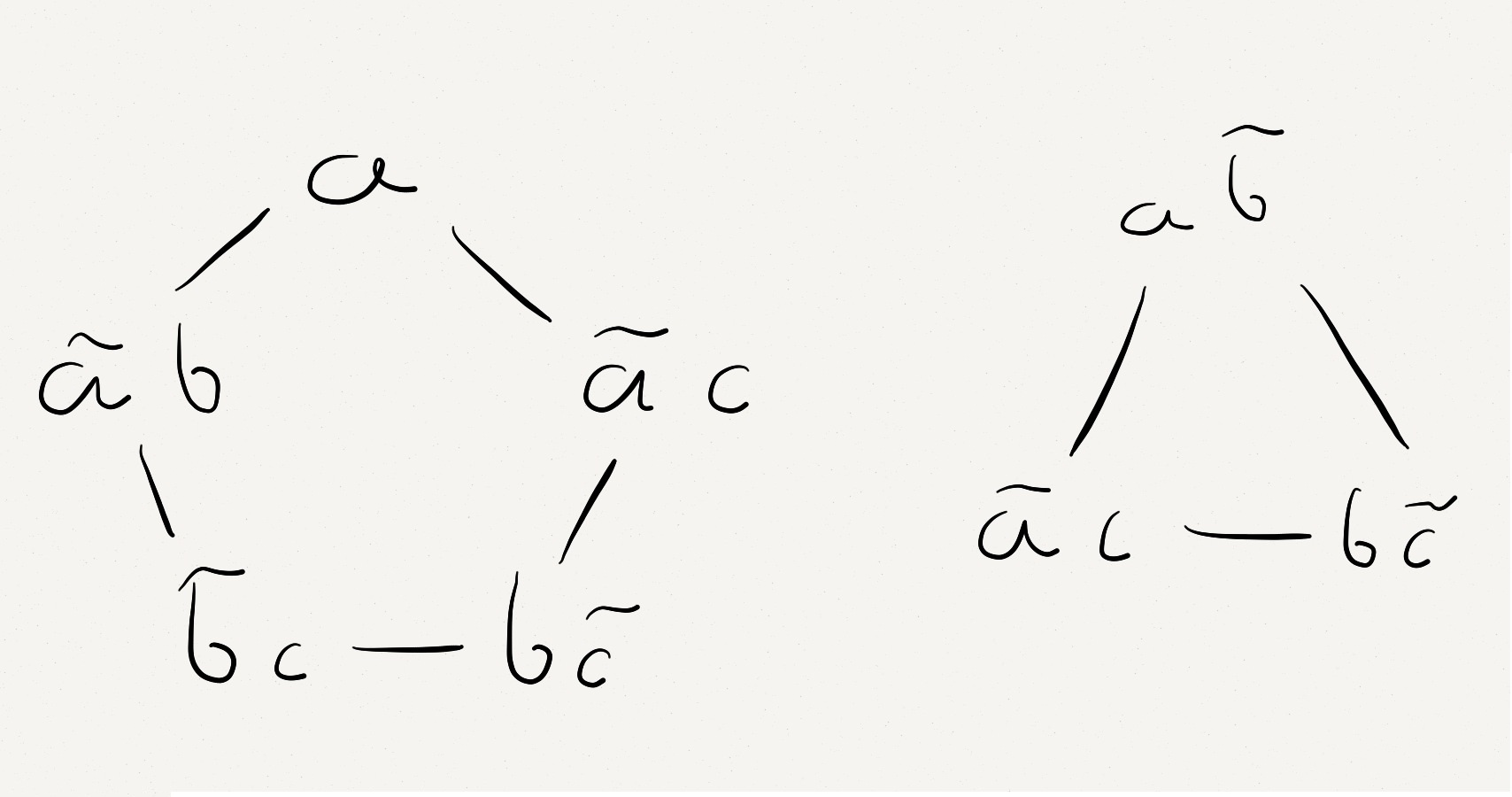}
\end{center}
\caption{\label{f2} Five cyclically exclusive events and three pairwise exclusive events. Edges denote exclusivity, not comeasurability.}
\end{figure}
Let us choose the following five cyclically exclusive events  $$a,~\tilde{a}b,~\tilde{b}c,~b\tilde{c},~\tilde{a}c,$$ whose exclusivity relationship is depicted in Fig. \ref{f2}. For instance, our non-contextuality assumption implies that if $A$ is assigned to be transmitted ($a$) then $A$ could not have been assigned to be reflected while $B$ was assigned to be transmitted ($\tilde{a}b$), hence $a$ and $\tilde{a}b$ must be exclusive. 
Hong-Ou-Mandel \cite{HOM} showed that photons bunch, that is, they exit together thru either one of the BS's output ports. On the other hand, if there is a single photon in one of the BS's input ports and the vacuum in the other, the photon exits thru either of the BS's output port with equal probability.  This implies that the conditional probabilities of the five events are $1/2$ and thus their sum is $5/2>\sqrt{5}$.

This result is contradictory to the recent proof by Cabello \cite{C2} that KCBS inequality violation cannot be greater than $\sqrt{5}$ in any no-disturbance theory with the exclusiveness as described in this paper. In his result the events used to construct the KCBS inequality are mutually exclusive but are not comeasurable, just like the ones introduced in this note. Moreover, one has to take into account two KCBS experiments performed in distant laboratories. Cabello assumes that probabilities of five pairwise exclusive events sum up to one, however as pointed by Henson \cite{IC2}, this does not have to be true. Here we show a system for which Cabello's assumption does not hold. In our case, if one considered an additional experiment on three photons performed in a distant laboratory and studied the five pairwise exclusive events corresponding to the ones discussed by Cabello, one would find that the sum of these events would reach $5/4$ (for the details of the construction of the events see Ref. \cite{C2}).  

Let us come back to the same setup of three photons in the fibers $A$, $B$, $C$ and a single $BS$, but this time we consider only three events $$a\tilde{b},~b\tilde{c},~c\tilde{a}.$$ Despite the fact that they are pairwise exclusive, quantum theory allows us to assign probability $1/2$ to each event and their sum is clearly equal to $3/2$. This is another example showing that probabilities of pairwise exclusive events do not have to sum up to one \cite{C2}. Furthermore, a non-contextual assignament of outcomes to these events would lead to an uperbound of 1 on the sum of their probabilities.  It implies that 3 events constitute the minimal set needed to reveal contextuality in a quantum mechanical system. 

To sum up, we demonstrated that bosonic nature cannot be described by a non-contextual hidden variable theory and found a simple physical system for which the sum of probabilities of pairwise exclusive events does not sum up to one. This implies that the KCBS-like scenario can give greater violation than it has been shown before. Our results can have implications on the role of quantum contextuality as a resource for quantum information processing \cite{prep}.  

Interestingly, Hong-Ou-Mandel could have performed the first test of contextuality already in 1987. The course of events forced us to wait yet another thirty four years to see Kochen and Specker's ideas materialize on the optical table \cite{Lapk}.

\end{document}